\documentclass[showpacs,amssymb,amsmath,aps,prl,reprint,groupaddress]{revtex4-1}
\usepackage{graphicx}
\usepackage{bm}
\usepackage{dsfont}
\usepackage{enumitem}
\usepackage[utf8]{inputenc}
\usepackage[T1]{fontenc}
\usepackage[english]{babel}

\begin{document}
\title{The Epidemic-Driven Collapse in a System with Limited Economic Resource}
\author{I.\,S.~Gandzha}
\email{gandzha@iop.kiev.ua}
\author{O.\,V.~Kliushnichenko}
\email{kliushnychenko@iop.kiev.ua}
\author{S.\,P.~Lukyanets}
\email{lukyan@iop.kiev.ua}
\affiliation{Institute of Physics, Nat.~Acad.~of~Sci.~of Ukraine, Prosp.~Nauky 46, Kyiv 03028, Ukraine}

\begin{abstract}
We consider a possibility of socioeconomic collapse caused by the spread of epidemic in a basic dynamical model with negative feedback between the infected population size and a formal collective economic resource. The epidemic-resource coupling is supposed to be of activation type, with the recovery rate governed by the Arrhenius-like law and resource playing the role of temperature. Such a coupling can result in the collapsing effect opposite to thermal explosion because of the limited resource. In this case, the system can no longer stabilize and return to the stable pre- or post-epidemic states. We demonstrate that such a collapse can partially be mitigated by means of a negative resource or debt.
\end{abstract}
\maketitle

Systemic shocks like the outbreak of epidemics and contagion spreading inevitably lead to negative socioeconomic outcomes \cite{SystemicShock_2020}. A dramatic example is the spread of COVID-19 that had a domino effect on both the social and economic levels. Different countries and governments resorted to different mitigation strategies and quarantine measures~\cite{Anderson_Lancet_2020-03}. Countries with a higher resource level (economic or financial) could use stricter quarantine measures, while for countries with a lower resource the use of such measures led to the economic collapse, at least for a number of industries and/or social groups. The problem of strategy selection reduces to problems of optimal control theory for feedback systems \cite{Economics_2014,Optimization_SIS_2017,Percolation_2017} or to the theory of games in a more general case \cite{Bauch_Earn_2004}. The use of one or another action strategy reduces to the classical problem of choice \cite{Economics}, i.e., to the definition of the sacrifice, when the salvation of someone or something is only possible at the expense of the other one.

To describe the socioeconomic interplay, the equations describing the spreading dynamics need to be coupled with equations for the dynamics of some formal economic resource \cite{Sugiarto_PRL_2017}. The mechanism of coupling between the spreading process and resource is of pivotal importance in this case. The resource's influence on the spreading process can naturally be taken into account by means of the resource-dependent recovery rate \cite{Swiss_2015,China_PRE_2019}, which is one of the basic parameters in many spreading models \cite{RevModPhys_2015}. In particular, it can approximately be fitted from empirical data as a function of the ratio between the infected population size and the average amount of resource devoted to infected individuals \cite{China_PRE_2019}. Such an epidemic-resource coupling can result in various critical and catastrophic phenomena, phase transitions, and multiphase behaviors \cite{China_PR_2018,China_2018}. On the other hand, the influence of the spreading process on the resource (or budget) can be taken into account in different ways, depending on the economic model adopted. Such an economic model can imply either the direct load on the budget (depending on the infected population size) or indirect mechanisms like taxes, etc.

The simplest example of the epidemic-resource coupling has been demonstrated for the basic susceptible-infected-susceptible (SIS) epidemic model, where the recovery rate was set dependent on the resource (budget) availability \cite{Swiss_2015}. A sufficiently wide class of model coupling functions was introduced to take into account the influence of the budget on the recovery rate. The counter effect on the budget was of direct and almost reciprocal character described by the same model function. The epidemic was shown to spiral out of control into ``explosive'' spread if the cost of recovery was above some critical cost. The similar explosive epidemic spreading can be observed in the case of connectivity disruption in networks \cite{Swiss_PRE_2016}. The spread of concepts, memes, hashtags as well as online rumor cascades can also be explosive~\cite{PRL_2020}.

In this work we consider a possibility for the coupling mechanism in the epidemic-resource system to be of activation type, with the recovery rate governed by the Arrhenius-like law. This idea stems from the fact that the recovery rate is generally determined by the quality of provision with medical services and food, apart from the individual peculiarities of the given member of population. The quickest recovery depends on the cost of medical services and the bare subsistence level of consumption ($E$) and the availability of the collective resource ($\rho$). Since the cost of services is fixed, the service is terminated if there is no sufficient resource ($\rho\ll E$). In other words, the parameter $E$ serves as the height of some energy barrier peculiar to the given system. Therefore, the recovery rate can be supposed to have an activation-type dependence, $\sim\exp(-E/\rho)$, similar to the temperature dependence of activation processes with activation energy $E$. In physical chemistry, such a dependence is known as the Arrhenius law \cite{Laidler,Stiller}.

The activation mechanism implies that the system can exhibit the so-called explosive (or catastrophic) instability \cite{PRL_1966}. For example, when a chemical reaction occurs with the release of heat and has an activation character, it goes faster at higher temperatures. This leads to yet greater temperature increase and ultimately to a thermal explosion, which is described in the framework of the Zel'dovich-Frank-Kamenetskii theory \cite{Zeldovich_Frank,Smirnov,Novozh_2018}.

The fight against the epidemic involves similar catastrophic processes. The spread of epidemic and the associated quarantine measures result in the reduction of the production of the collective resource $\rho$. When the resource is depleted, the quality of medical services drops and the recovery rate goes down. As a result, the number of active members in the population decreases. This, in turn, leads to a further decline in the collective resource production, with the level of income needed for the basic survival being lower and lower. Such a scenario finally results in the complete collapse of the system---the effect opposite to thermal explosion.

The purpose of this work is to demonstrate a possibility of (i) such an epidemic-driven collapse where the epidemic-resource coupling is supposed to be of activation type and (ii) partial mitigation of this scenario by adopting an economic model with negative resource (debt) and the so-called negative income tax \cite{NIT_Chicago,NIT_2020}.

To describe the collapse dynamics, we resort to a simple SIS-like model supplemented with the resource (budget) equation implying the indirect influence of the epidemic on the resource:
\begin{equation}\label{eq:SIS}
\begin{split}
\partial_t s &= - \beta\, s\,\bigl(1-s\bigr) + \gamma(\rho)\,\bigl(1-s\bigr),\\
\partial_{t} \rho &= G\,s-\Gamma\rho-\Lambda.
\end{split}
\end{equation}
The operator $\partial_t$ stands for the derivative with respect to time $t$. Here $s$ is the number density of susceptible individuals (active population) which are infected at some transmission rate $\beta$ defined as a product of the contact rate and the probability that a contact of an infected individual with a susceptible individual results in transmission. The infected individuals with number density $i=1-s$ recover and become susceptible again with resource-dependent recovery rate
\begin{equation}\label{eq:gamma}
\gamma(\rho)\simeq \gamma_0 \,\exp(-E/\rho).
\end{equation}
The recovery process is governed by the general economic situation described by some integral activation parameter $E$ which reflects the cost of medical and other essential life services as well as the bare subsistence level of consumption.

The function $\rho$ represents a collective economic resource or budget. The production of this resource per unit time is proportional to the number density of working (active) individuals, $s$. The parameter $G$ formalizes the resource volume generated by them per unit time. The second term, $\Gamma\,\rho$, formally describes the collective expenses or taxes. Roughly speaking, the expenses are assumed to be proportional to earnings. Thus, the coefficient $\Gamma$ represents the resource consumption rate. The parameter $\Lambda$ represents the overall fixed expenses necessary for keeping some infrastructure (e.g., amortization, municipal services, etc.). Our resource balance equation is in accord with an equation for the budget constraint in the macroeconomic model considered in Ref.~\cite{Macroeconomic_2020}. In contrast to the budget equation proposed in Ref.~\cite{Swiss_2015}, we consider the indirect influence of the epidemic on the budget via taxes, collective expenses, or infrastructure.

The initial conditions at $t=0$ are taken as
\begin{equation}
s(0) = 1 - i_0,\quad i(0) = i_0,\quad \rho(0) = \rho_0,
\end{equation}
$i_0$ being the initial number density of infected individuals.

In the case of unlimited resource ($E\ll\rho$), the equation for $s$ reduces to the basic SIS model, whose solutions are well studied~\cite{BookSpringer2015,BookSpringer2019}. This model is due to Kermack and McKendrick \cite{KermackMcKendrick}; it is also known as the Schl\"{o}gl~I model describing autocatalytic chemical reactions \cite{Schlogl_1972}. The SIS model can also be derived as the mean-field approximation to more general network models \cite{RevModPhys_2015,Swiss_PRL_2017}.

Let us investigate the effect of nonzero activation parameter $E$ (activation energy) on the coupled epidemic-resource dynamics described by system~(\ref{eq:SIS}). We first consider the simplest case with no infrastructure expenses ($\Lambda=0$). The case of nonzero $\Lambda$ is considered afterwards.

{\it No infrastructure expenses}.---As we understand, if there are no systemic shocks like the epidemic, there exists a stationary equilibrium state with $\rho = \rho^{(0)} = \mathrm{const}$. It is called the disease-free equilibrium and is given by a trivial stationary solution to Eqs.~(\ref{eq:SIS}), namely, $s^{(0)}=1$, $\rho^{(0)}=G/\,\Gamma$.

Under systemic shock conditions like epidemic, the system can go out from the disease-free equilibrium, with resource decreasing. Indeed, apart from the disease-free trivial solution $\rho^{(0)}$, another stationary solution to the equation for $\rho$ is given by $\rho^{*}=G\,s^{*}/\,\Gamma$, where $s^{*}$ is given by the following transcendental equation:
\begin{equation}\label{eq:SIS_trans}
s^{*}\,\ln\left(\mathcal{R}_0 s^{*}\right)=-\mathcal{E},
\end{equation}
provided that $s^{*}>0$. The dimensionless parameter $\mathcal{R}_0=\beta/\,\gamma_0$ is well known as the basic reproduction number. It defines the average number of transmissions one infected individual makes in the entire susceptible compartment during the entire time of being infected. The dimensionless parameter $\mathcal{E}=E/\,\rho^{(0)}$ is the activation energy normalized by the stationary resource value $\rho^{(0)}$.

\begin{figure}[t]
\includegraphics[width=0.75\columnwidth]{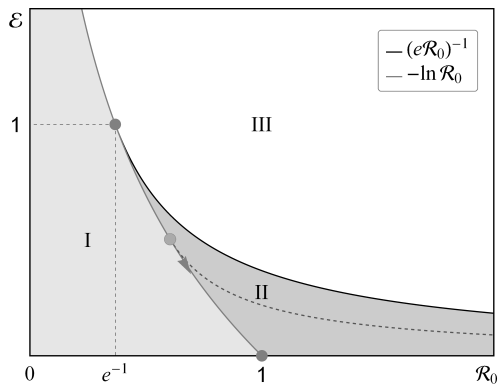}
\caption{\label{fig:phase}Phase diagram for the coupled epidemic-resource system described by Eqs.~(\ref{eq:SIS}) with $\Lambda=0$. Three states (phases) are possible: (I) disease-free equilibrium or collapse (subject to initial conditions), (II) endemic equilibrium or collapse (subject to initial conditions), and (III) collapse. The dashed curve is the position of the boundary between phases II and III (solid black curve) in the case of nonzero $\Lambda$ ($s_\Lambda = \Lambda/\,G = 0.4$).}
\end{figure}

\begin{figure*}[!]
\includegraphics[width=\textwidth]{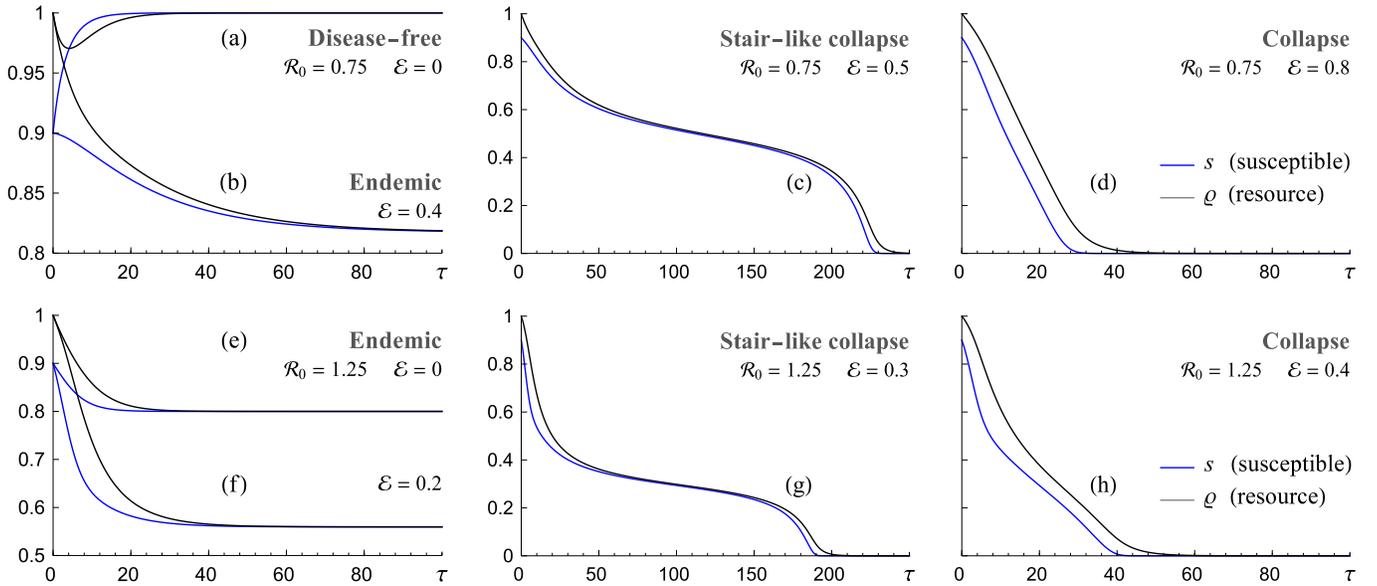}
\caption{\label{fig:SIS1}The number density of susceptible individuals and normalized resource function $\varrho=\rho/\rho^{(0)}$ versus dimensionless time $\tau = \gamma_0\,t$ in the coupled epidemic-resource system described by Eqs.~(\ref{eq:SIS}) with $\Gamma/\gamma_0=0.2$, $\mathcal{R}_0=0.75$ or $\mathcal{R}_0=1.25$, and various normalized activation energies $\mathcal{E}$ in the case of zero infrastructure expenses ($\Lambda=0$). The initial conditions are $i_0=0.1$, $\varrho_0 = 1$. When $\mathcal{E}=0$, the system evolves to (a)~disease-free equilibrium at $\mathcal{R}_0<1$ and (e) endemic equilibrium at $\mathcal{R}_0>1$. When $\mathcal{E}>0$, the system evolves either to the endemic equilibrium (phase II in Fig.~\ref{fig:phase}) both for (b) $\mathcal{R}_0<1$ and (f) $\mathcal{R}_0>1$ or collapses (phase~III in Fig.~\ref{fig:phase}). When $\mathcal{E}$ is above the critical value $\mathcal{E}_c$ given by Eq.~(\ref{eq:E_c}) but still close to it, the system first tries to occupy the quasi-stationary endemic state (which no longer exists). This process can take quite a long time and then the system finally collapses both for (c) $\mathcal{R}_0<1$ and (g) $\mathcal{R}_0>1$ (the so-called stair-like collapse). At larger activation energies, the collapse is very fast with no intermediate quasi-stationary evolution both for (d) $\mathcal{R}_0<1$ and (h) $\mathcal{R}_0>1$.}
\end{figure*}

\begin{figure*}[!]
\includegraphics[width=\textwidth]{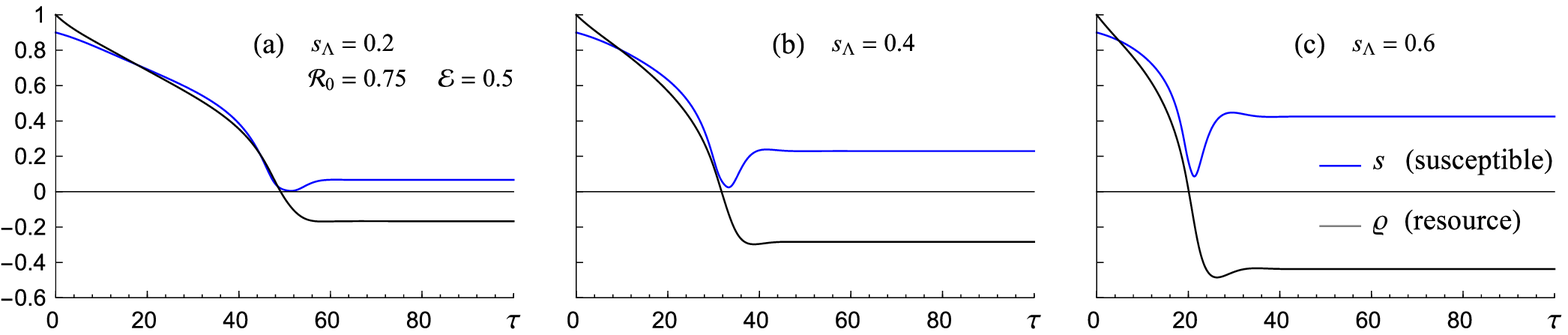}
\caption{\label{fig:SIS2}The number density of susceptible individuals and normalized resource function $\varrho=\rho/\rho^{(0)}$ versus dimensionless time $\tau = \gamma_0\,t$ in the case of nonzero infrastructure expenses ($\Lambda\ne0$) with all other parameters selected as in Fig.~\ref{fig:SIS1}(c). The collapse scenario is mitigated owing to negative resource (debt). The larger the infrastructure expenses $\Lambda$ ($s_\Lambda = \Lambda/\,G$), the greater is the debt required to keep the infrastructure operational and the larger is the number of active (recovered) individuals.}
\end{figure*}

When $\mathcal{E}=0$, Eq.~(\ref{eq:SIS_trans}) has one solution, $s^{*}=\mathcal{R}_0^{-1}$ (since $s^{*}\ne0$). It is stable at $\mathcal{R}_0 > 1$ and defines the endemic equilibrium point. Thus, as it is well-known for the case of unlimited resource $\mathcal{E}=0$, the disease-free equilibrium is stable when $\mathcal{R}_0\leqslant 1$, and there is no epidemic outbreak \cite{BookSpringer2015,BookSpringer2019}. When $\mathcal{R}_0 > 1$, the disease-free equilibrium is unstable, and the system evolves to the new equilibrium state $\{s^*,\rho^*\}$ called the endemic equilibrium.

When $\mathcal{E}>0$, there are several possible cases. For $0<\mathcal{E}<\mathcal{E}_c$, where
\begin{equation}\label{eq:E_c}
\mathcal{E}_c=\left(e\mathcal{R}_0\right)^{-1},
\end{equation}
Eq.~(\ref{eq:SIS_trans}) has two solutions: $s^{*}_1 > \mathcal{E}_c$ (which defines the endemic equilibrium point) and $0<s^{*}_2 < \mathcal{E}_c$ (which is always unstable). For $\mathcal{E}=\mathcal{E}_c$, there is one solution $s^{*}_{1,2} = \mathcal{E}_c$. Finally, there are no real solutions for $\mathcal{E}>\mathcal{E}_c$.

Except for the condition $0<\mathcal{E}\leqslant\mathcal{E}_c$, the endemic equilibrium point should also meet the requirement of $s^{*}< 1$, which effectively implies that $\mathcal{E}>\mathcal{E}_e$, where
\begin{equation}\label{eq:E_log}
\mathcal{E}_e=-\ln\mathcal{R}_0.
\end{equation}
The same relation can be obtained from the stability analysis of the disease-free stationary solution $s^{(0)}=1$. The  disease-free equilibrium is stable at $\mathcal{E}\leqslant \mathcal{E}_e$ and unstable at $\mathcal{E} > \mathcal{E}_e$.

In the case $\mathcal{E}>0$, system~(\ref{eq:SIS}) also possesses another stable stationary solution given by $s^{*}_c=0$, $\rho^{*}_c\rightarrow0$. This means that at any nonzero $\mathcal{E}$ and $\mathcal{R}_0$ there exist such initial conditions $\{i_0,\,\rho_0\}$ that the system collapses (\mbox{$s\rightarrow 0$}) because of resource depletion ($\rho\rightarrow0$). When $\mathcal{E}\leqslant \mathcal{E}_e$, the stationary point $s^{*}_c=0$ coexists with the disease-free equilibrium $s^{(0)}=1$. The system evolves to one of these two points, depending on initial conditions $\{i_0,\,\rho_0\}$. Similarly, the stationary point $s^{*}_c=0$ coexists with the endemic equilibrium $s^{*}$ when $\mathcal{E}_e<\mathcal{E}\leqslant \mathcal{E}_c$. Finally, the system collapses at any initial conditions when $\mathcal{E}> \mathcal{E}_c$.

Thus, relations (\ref{eq:E_c}) and (\ref{eq:E_log}) define two critical curves in the $(\mathcal{R}_0,\,\mathcal{E})$ plane which determine the evolution scenario for dynamical system (\ref{eq:SIS}). Depending on the values of parameters $\mathcal{R}_0$ and $\mathcal{E}$, the system can evolve into three possible states (phases): disease-free equilibrium, endemic equilibrium, or collapse. Figure \ref{fig:phase} shows the corresponding phase diagram.

The above analysis is supported by the results of numerical integration of Eqs.~(\ref{eq:SIS}) demonstrated in Fig.~\ref{fig:SIS1}.

When $\mathcal{E}=0$, the dynamics of system~(\ref{eq:SIS}) follows the basic SIS model. It evolves to the state of disease-free equilibrium at $\mathcal{R}_0\leqslant1$ [Fig.~\ref{fig:SIS1}(a)] and to the state of endemic equilibrium at $\mathcal{R}_0>1$ [Fig.~\ref{fig:SIS1}(e)].

When $\mathcal{E}>0$, some part of the resource is consumed, and the number of active (susceptible) individuals decreases [Fig.~\ref{fig:SIS1}(f)]. There is a critical value $\mathcal{E}_e$ defined by formula~(\ref{eq:E_log}) at which the system evolves to the endemic equilibrium even at $\mathcal{R}_0<1$ [Fig.~\ref{fig:SIS1}(b)]. This scenario is impossible in the basic SIS model. For the activation energies larger than the critical value $\mathcal{E}_c$ defined by formula~(\ref{eq:E_c}), the endemic equilibrium is no longer stable and the system collapses to the state $s^{*}_c=0$, $\rho^{*}_c\rightarrow0$ [Fig.~\ref{fig:SIS1}(c,d)]. This means that all the individuals become infected and there is no resource to reverse the epidemic back. The same scenario is observed in the case $\mathcal{R}_0>1$ [Fig.~\ref{fig:SIS1}(g,h)]. Note that the endemic equilibrium shown in Fig.~\ref{fig:SIS1}(b,f) coexists with the collapse point, the system's dynamics switching from endemic to collapse at small $\rho_0$ and large $i_0$.

When $\mathcal{E}$ is above the critical value $\mathcal{E}_c$ but still close to it, the system first tries to occupy the quasi-stationary endemic state (which no longer exists). This process can take quite a long time and then the system finally collapses [Fig.~\ref{fig:SIS1}(c,g)]. It resembles the well-known ``devil's staircase'' pattern \cite{Staircase}. At larger activation energies, the collapse is very fast with no intermediate quasi-stationary evolution [Fig.~\ref{fig:SIS1}(d,h)].

{\it Nonzero infrastructure expenses}.---Surprisingly but nonzero infrastructure expenses ($\Lambda>0$) can mitigate (or at least stabilize) the above-described hard collapse scenario. In fact, nonzero $\Lambda$ breaks the symmetry of the equation for $\rho$, so that the point $s^{*}_c=0$, $\rho^{*}_c\rightarrow0$ might no longer be its stationary solution. On the other hand, there exists another stationary point,
\begin{equation}\label{eq:rhostar}
\Gamma \rho^{*}= Gs^{*}-\Lambda,
\end{equation}
where the resource value $\rho^*$ can become negative at sufficiently small but positive $s^{*}$. This means that at certain conditions there is no sufficient collective resource to sustain the infrastructure expenses, and some of these expenses need to be financed through a debt. In this case, the second term in the equation for $\rho$ becomes positive indicating that the resource is no longer consumed but is ``pumped'' back into the system in the form of external subsidies or the so-called negative income tax \cite{NIT_Chicago,NIT_2020}.

Such an economic model naturally follows from the form of our equation for the resource dynamics. To keep the recovery rate finite in the case of negative resource, it is necessary to rewrite relation~(\ref{eq:gamma}) in terms of the resource's absolute value, namely,
\begin{equation}
\gamma(\rho)\mapsto\gamma(|\rho|)= \gamma_0 \,\exp(-E/\,|\rho|),
\end{equation}
with asymptotic value $\gamma(\rho)=0$ at $\rho=0$.

Now, the disease-free trivial stationary solution to Eqs.~(\ref{eq:SIS}) is $s^{(0)}=1$, $\rho^{(0)}=(G-\Lambda)/\,\Gamma$. In what follows, we will restrict our attention to the case $\Lambda<G$.

The nontrivial stationary number density $s^{*}$ of active population is given by the transcendental equation
\begin{equation}\label{eq:SIS_trans_Lambda}
|s^{*}-s_\Lambda|\,\ln\left(\mathcal{R}_0 s^{*}\right)=-\mathcal{E}\left(1-s_\Lambda\right),
\end{equation}
where the parameter $s_\Lambda = \Lambda/\,G<1$ defines the minimum number of active individuals required to keep the infrastructure operational. This equation can be analyzed similarly to Eq.~(\ref{eq:SIS_trans}), with the critical activation energy now having the form $\mathcal{E}_c=\varepsilon\left(\mathcal{R}_0(1-s_\Lambda)\right)^{-1}$, where $\varepsilon$ is the absolute value of the local minimum of the transcendental function of $s^*$ in the left-hand side of Eq.~(\ref{eq:SIS_trans_Lambda}). In particular, $\varepsilon=e^{-1}$ for $s_\Lambda=0$ [see Eq. (\ref{eq:E_c})] and $\varepsilon=0$ for $s_\Lambda=\mathcal{R}_0^{-1}$. As $s_\Lambda$ increases from 0 to $\mathcal{R}_0^{-1}$, the triple point in the phase diagram shown in Fig.~\ref{fig:phase} slides down across the logarithmic curve $\mathcal{E}=\mathcal{E}_e$ until it reaches the point $\mathcal{E}=0$.

Note that, in contrast to the case $\Lambda=0$, Eq.~(\ref{eq:SIS_trans_Lambda}) has one additional solution, $s_\rho^{*}<s_\Lambda$, that exists for any \mbox{$\mathcal{E}>0$}. It is always stable, and the corresponding stationary resource value is always negative. This solution is the direct counterpart to the collapse point $s_c^{*}=0$ existing in the case $\Lambda=0$. Thus, nonzero $\Lambda$ serves as a mitigating factor to the collapse scenario, with the system stabilization achieved owing to negative resource (debt). Figure~\ref{fig:SIS2} demonstrates such a mitigated collapse scenario for the same set of parameters as in Fig.~\ref{fig:SIS1}(c). The number density of active population bounces from a horizontal axis close to $s=0$ and stabilizes at $s=s_\rho^{*}$, with resource passing through the zero point and stabilizing at the negative value given by Eq.~(\ref{eq:rhostar}). The larger the parameter $s_\Lambda$, the greater is the debt required to keep the infrastructure operational and the larger is the number of active (recovered) individuals.

A simple model for the socioeconomic system considered here is based on the activation-type mechanism of the epidemic-resource coupling [see Eq.~(\ref{eq:gamma})]. Such a coupling mechanism naturally results in the collapsing effect opposite to the well-known thermal explosion. The activation parameter $E$ characterizes the minimum amount of the consumed resource needed for the survival of a community or a particular individual, therefore implying the existence of some ``energy'' barrier for their survival. Similar mechanisms are likely to be peculiar to other population systems as well.

In this work, we demonstrated that in the case of limited economic resource there exists a certain critical point at which the system collapses at any initial conditions and can no longer stabilize and return to the stable pre-epidemic or post-epidemic state. Such a scenario is possible even when the basic reproduction number $\mathcal{R}_0$ is smaller than unity, in contrast to the standard epidemic models, where the epidemic can spread only at $\mathcal{R}_0>1$ \cite{BookSpringer2015,BookSpringer2019}. We also demonstrated that the system’s collapse can partially be mitigated by the overexpenditure of the budget interpreted as a negative resource or debt and by adopting an economic model with negative income tax \cite{NIT_Chicago,NIT_2020} securing the necessary number of active individuals to keep the minimum infrastructure operational. These results provide a clear illustration to the possible outcomes of systemic shocks in the global pandemic scenario.

\vspace{6pt}

O.K. was partially supported by a grant for research groups of young scientists from the National Academy of Science of Ukraine (Project No. 0120U100155). We thank Prof. B.I. Lev for fruitful discussions.

\selectlanguage{english}

\end{document}